\documentclass[prb,twocolumn,superscriptaddress,showpacs,showkeys]{revtex4}

\usepackage{amssymb,amsfonts}
\usepackage{graphicx}
\usepackage{epsfig}
\newcommand{\be}{\begin{equation}}\newcommand{\ee}{\end{equation}}
\newcommand{\bea}{\begin{eqnarray}}
\newcommand{\eea}{\end{eqnarray}}
\newcommand{\beq}{\begin{eqnarray}}
\newcommand{\eeq}{\end{eqnarray}}

\newcommand{\al}{\alpha}

\begin{document}

\title{Casimir interaction between a perfect
conductor and graphene described by the Dirac model}
\affiliation{Institut f\"ur Theoretische Physik, Universit\"at
Leipzig, Vor dem Hospitaltore 1, 04103 Leipzig, Germany}
\affiliation{Instituto de F\'isica, Universidade de S\~ao Paulo,
Caixa Postal 66318 CEP 05314-970, S\~ao Paulo, S.P., Brazil}
\affiliation{CMCC, Universidade Federal do ABC, Santo Andr\'e,
S.P., Brazil}
\affiliation{Department of Theoretical Physics, St.\ Petersburg State University, St.\ Petersburg,
Russia}
\author{M.~Bordag}
\affiliation{Institut f\"ur Theoretische Physik, Universit\"at Leipzig, Vor dem Hospitaltore 1,
04103 Leipzig, Germany}
\author{I.~V.~Fialkovsky}
\affiliation{Instituto de F\'isica, Universidade de S\~ao Paulo, Caixa Postal 66318 CEP 05314-970,
S\~ao Paulo, S.P., Brazil}
\affiliation{Department of Theoretical Physics, St.\ Petersburg State University, St.\ Petersburg,
Russia}
\author{D.~M.~Gitman}
\affiliation{Instituto de F\'isica, Universidade de S\~ao Paulo, Caixa Postal 66318 CEP 05314-970,
S\~ao Paulo, S.P., Brazil}
\author{D.~V.~Vassilevich}
\affiliation{CMCC, Universidade Federal do ABC, Santo Andr\'e, S.P., Brazil}
\affiliation{Department of Theoretical Physics, St.\ Petersburg State University, St.\ Petersburg,
Russia}
\begin{abstract}
We adopt the Dirac model for graphene and calculate the Casimir
interaction energy   between a plane suspended graphene sample and
a parallel plane perfect conductor. This is done in two ways.
First, we use the Quantum Field Theory (QFT) approach and evaluate
the leading order diagram in a theory with $2+1$ dimensional
fermions interacting with $3+1$ dimensional photons. Next, we
consider an effective theory for the electromagnetic field with
matching conditions induced by quantum quasi-particles in graphene.
The first approach turns out to be the
leading order in the coupling constant of the second one. The
Casimir interaction for this system appears to be rather weak. It
exhibits a strong dependence on the mass of the quasi-particles in
graphene.
\end{abstract}
\maketitle
\section{Introduction}
Graphene is a (quasi) two dimensional hexagonal lattice of carbon
atoms. At present, it belongs to the most interesting materials in
solid state physics in view of its exceptional properties and its
potential applications in nano technology (see reviews, Refs.~\cite{gra-rev,RMP}). At small separations, nearly down to contact,
the interaction between a graphene sample and any
solid body (dielectric, conductor or another graphene) is due to the van der Waals
and, at larger separations, Casimir forces (see, e.g.
Refs.~\cite{Casimir1,Casimir2}). The latter are the subject of the present
paper. We consider the same geometry as in the original Casimir
effect (2 parallel planes) with one plane being
graphene and the other one -- ideal conductor. This setup was considered in
Refs.~\cite{BV,Bordag:2005by,BGKM} using a hydrodynamical model for the
electrons in graphene following Refs.~\cite{Fetter73,BIII}.
Later it became clear that this model does not describe the electronic
properties specific to this novel material.
Here we use a realistic and well-tested model
where the quasi-particles in graphene are considered to be fermions subject to the Dirac
equation which models their linear dispersion law.
Details of this model can be found in Refs.~\cite{gra-rev,DiracModel}.

The result presented below is the first calculation of the Casimir
interaction of graphene made within a reasonable theoretical model.
Although the hydrodynamical (plasma) model is not applicable to graphene,
it works well for some other materials, and it will be used as a theoretical
reference point to compare our results.

We like to mention some related works.
The Casimir-like interaction between adatoms due to fermionic modes
{\textit{inside}} graphene was studied in Ref.~\cite{adatoms}. The graphene-metal
interaction at separations $2$--$4${\AA}
(much smaller than the Casimir distances) was investigated in Refs.~\cite{2A}.

Let us formulate the model.
The electronic properties of graphene are well described by the above mentioned
Dirac model \cite{DiracModel}.
It incorporates the most essential and well-established properties of
the quasi-particles' spectrum: the linearity, a very small mass gap (if any),
and a characteristic propagation velocity which is $300$
times smaller than the speed of light. The model deals, therefore,
with light fermions in $2+1$ dimensions
(confined to the surface of graphene) with the action
\begin{equation}
S_{\rm D}=\int d^3x \bar\psi (\tilde \gamma^l
(i\partial_l-eA_l)-m)\psi, \label{Di}
\end{equation}
where $l=0,1,2$. The matrices $\tilde\gamma^l$ are rescaled,
$\tilde\gamma^0\equiv\gamma^0,\ \tilde\gamma^{1,2}\equiv
v_F\gamma^{1,2},\ \gamma_0^2=-(\gamma^1)^2=-(\gamma^2)^2=1$.
$v_F$ is the Fermi velocity. In our units, $\hbar=c=1$, $v_F\simeq
(300)^{-1}$.  The gamma matrices are taken in the form of a direct
sum of two inequivalent representations (differing by an overall
sign). There is an additional (`valley') degeneracy in graphene, so that we have $4$ two-component
spinors or $2$ four-component spinors in graphene.
The value of the mass gap parameter $m$ and mechanisms of its generation
are under discussion \cite{Appelquist:1986fd,massgap1,Gusynin,Pyatkovskiy}.
The upper limit on $m$ is about $0.1$eV, but probably is much smaller.
Due to this smallness of the mass, the quasi-particles exhibit a relativistic
behavior at rather small energies, which makes QFT a more adequate language to describe graphene
than Quantum Mechanics.
As we shall see below, the Casimir force is very sensitive
to the value of $m$, and thus the Casimir experiments may be used
to estimate~it.

The propagation of photons in the ambient $3+1$
dimensional space is described by the Maxwell action
\begin{equation}
S_{\rm M}=-\frac 14 \int d^4x F_{\mu\nu}F^{\mu\nu},\quad
\mu,\nu=0,1,2,3. \label{SEM}
\end{equation}
The coupling constant is normalized according to
$e^2/(4\pi)=\alpha\simeq 1/137$. The Dirac model with quantized
fermionic quasi-particle excitations and classical electromagnetic
modes describes rather well the optical properties of graphene. By
construction, this model should work below the energy scale of
about $1$eV, but even above this limit the absorption of light by
suspended layers of graphene is reproduced with a high precision
\cite{Nair}.

In the following we calculate the Casimir force between a flat
suspended monolayer graphene sample and a parallel flat perfect
conductor. We shall suppose that the graphene sample occupies the
plane $x^3=a>0$, and the conductor corresponds to $x^3=0$.\\

\section{QFT approach}
One possible way to calculate the Casimir energy in the system in
question is to evaluate the effective action $\Gamma$ in a quantum
field theory described by the classical action $S_{\rm D}+S_{\rm
M}$. Since the background is static (the positions of the surfaces
do not depend on time), the energy density per unit area of the
surfaces is ${\mathcal {E}}=-\Gamma/(TS)$, where, because of the
translation invariance, one has to divide the effective action
$\Gamma$ by the (temporarily
introduced) time interval $T$ and the area of the surface $S$. At
the leading order in the fine structure constant $\alpha$ we have
\begin{equation}
\mathcal{E}_1= -\frac 1{TS}
\raisebox{-3.75mm}{\psfig{figure=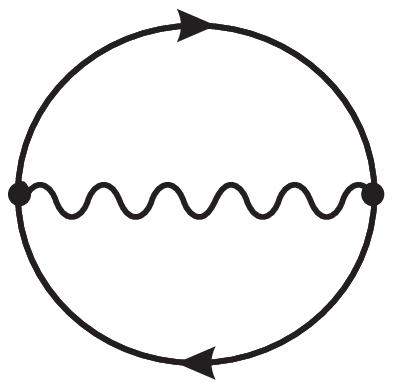,height=.4in}},
\label{E1}
\end{equation}
where the solid line denotes the fermion propagator in $2+1$ dimensions
(i.e., inside the graphene sample), and the wavy line is the photon propagator
in the ambient $3+1$ dimensional space subject to the perfect conductor
boundary conditions
\begin{equation}
A_0\vert_{x^3=0}=A_1\vert_{x^3=0}=A_2\vert_{x^3=0}=\partial_3A_3\vert_{x^3=0}=0.
\label{condbc}
\end{equation}
We  use the Feynman gauge such that no contribution from ghosts
appears. For details on diagrammatic notations see Ref.~\cite{Ramon}.

The calculation of (\ref{E1})
is rather similar to that of the radiative corrections to the Casimir energy
in Ref.~\cite{radicor}. The important difference to those works is that they considered the
fermions propagating in the ambient space as a correction to the
Casimir force between two perfect conductors, which appeared to be very
small. In that case the diagram (\ref{E1}) represents
only the $O(\al)$ correction to the Casimir energy.  In our case, the diagram (\ref{E1}),
still being $O(\al)$, represents the leading order effect.

 The fermion loop in $2+1$ dimensions has already been
calculated in a number of papers
\cite{Appelquist:1986fd,Gusynin,Pyatkovskiy}. This result is most
conveniently expressed in terms of the effective action for
fermions in the presence of an external electromagnetic field,
$S_{\rm eff}(A)=-i \ln\det ((\tilde\gamma^l(i\partial_l-eA_l)-m)$. To
the quadratic order in $A$ the effective action reads
\begin{eqnarray}
&&S_{\rm eff}(A)= A \ \raisebox{-3.75mm}{\psfig{figure=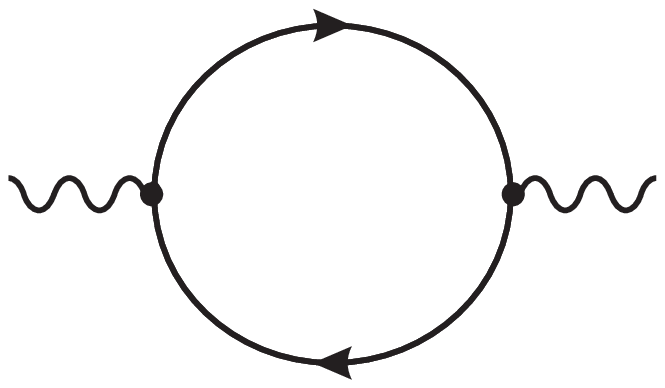,height=.4in}}\  A\nonumber\\
&&\qquad\  = \frac 12 \int \frac{d^3p}{(2\pi)^3} A_j(p) \Pi^{jl}(p)A_l(p),\label{Seff}
\end{eqnarray}
where
\begin{eqnarray}
 &&\Pi^{mn}
        = \frac {\alpha\Phi(\tilde p)}{v_F^2} \,
             \eta^{m}_{j}\left(g^{jl}-\frac{\tilde p^j\tilde p^l}{\tilde p^2}\right)\eta_l^n\,,
\label{Pmn}\\
&&\Phi({p})=
        N\frac{2 m \tilde p -(\tilde p^2+4m^2){\rm arctanh }({\tilde p}/{2m})}{2\tilde
        p}
\label{Phi}
\end{eqnarray}
is the polarization tensor in the lowest, one loop, order.
Here
$\eta^{m}_j={\rm diag}(1,v_F,v_F)$, and $\tilde p$ denotes the
rescaled momenta $\tilde p_j=\eta_j^k p_k$. $N$ is the number of
two-component fermion species, $N=4$ for graphene. Note, that in
chosen above representation of gamma
matrices the parity-odd parts of the polarization tensor $\Pi$ are canceled
between contributions of various fermion species. If they are not,
due to external magnetic field or for some other reason, this may
be measured by studying the polarization rotation of light passing
through suspended graphene \cite{IFDV}.
It is also essential to notice that both polarization operator $\Pi$
(\ref{Pmn}) and the diagram in (\ref{E1}) do not posses infra--red
divergencies in the  $m=0$ limit.

To calculate the diagram (\ref{E1}) we only need to couple the kernel (\ref{Pmn}) to the
photon propagator and integrate over the photon momenta. Symbolically,
\begin{equation}\label{Gamma}
   \Gamma=\frac i2 {\rm Tr}(\Pi D),
\end{equation}
where ${\rm Tr}$ is the functional trace, and $D$ is the
photon propagator. In the Feynman gauge the propagator
$D_{\mu\nu}$ is diagonal. Since $\Pi^{ij}$ by construction does not have
components along the $x^3$ direction, we are only interested in
the $D_{ij}$ part of the propagator. According to (\ref{condbc}), these
components satisfy Dirichlet boundary conditions, and one can
write
\begin{equation}
    D_{ij}(x;y)=g_{ij} (D_0(x-y)-D_0(x-y_R))\,,\label{Dij}
\end{equation}
where $D_0(x-y)$ is the standard (Feynman) propagator of a free massless scalar field in $3+1$ dimensions, and the
coordinate $y_R$ is reflected at the conductor surface, $y_R^3=-y^3$.
The full effective action $\Gamma$ thus reads
\begin{equation}
    \Gamma=\frac i2 \int d^3x\, d^3y\, \Pi_j^j(x,y) [ D_0(x-y,0)-D_0(x-y,2a)] \,,\label{Gam}
\end{equation}
where both $x$ and $y$ lay on the surface of graphene.
After making the Fourier
transform in the directions parallel to the surfaces, one can write $D_0$ as
\begin{equation}
    D_0(x,y)=\int \frac{d^3p}{(2\pi)^3}\, e^{ip_j(x^j-y^j)}\,
    D(p,x^3-y^3).\label{D0xy}
\end{equation}
For the Euclidean $3$-momenta, i.e., after the Wick rotation
$p\to p_E=(p_4,p_1,p_2)$, $p_4=ip_0$, its explicit form reads
\be
    D(p_E,x^3-y^3)=\frac{e^{-p_{\|} |x^3-y^3|}}{2p_\|},\qquad p_\| \equiv |p_E|.
\label{DpE}
\ee
The term with $D_0(x-y,0)$ on the right hand side of (\ref{Gam}) does not depend
on $a$ and will be neglected (as it does not contribute to the Casimir force).
The remaining terms in $\Gamma$ are non-divergent. After the Fourier transformation
and the Wick rotation we obtain
\begin{eqnarray}
&&\mathcal{E}_1\equiv
    -\frac\Gamma {TS}=-\frac 14 \int\frac{d^3p_E}{(2\pi)^3}\,
\frac{\Pi_j^j(p_E)}{p_\|}\, e^{-2ap_\|} \nonumber\\
 &&\qquad   =-\frac 14 \int\frac{d^3p_E}{(2\pi)^3}\,
        \frac{\al (p_\|^2+\tilde p_\|^2)\Phi(p_E)}{p_\|^{\phantom{2}} \tilde p_\|^2}\, e^{-2ap_\|}\,. \label{E11}
\end{eqnarray}
where we expanded  $\Pi_j^j(p_E)$ explicitly  with help of (\ref{Pmn}).

\section{Lifshitz formula approach}
One can also adopt another point of view on the Casimir effect for this system
and consider an effective theory of the electromagnetic field described
by the action $S_{\rm M}+S_{\rm eff}$ subject to the conductor boundary
conditions (\ref{condbc}) at $x^3=0$. Away from the surfaces, the photons
propagate freely. They are reflected at the surface of the conductor $x^3=0$.
At the surface of graphene, the Maxwell equations receive a singular contribution
\begin{equation}
\partial_\mu F^{\mu\nu}+\delta(x^3-a)\Pi^{\nu\rho}A_\rho =0\label{Meq}
\end{equation}
following from $S_{\rm eff}$. Here we extended $\Pi$ to a $4\times 4$ matrix by setting
$\Pi^{3j}=\Pi^{j3}=0$. This singular contribution is equivalent to imposing the matching
conditions
\begin{eqnarray}
&& A_\mu\vert_{x^3=a+0}=A_\mu\vert_{x^3=a-0},\nonumber\\
&& (\partial_3 A_\mu)\vert_{x^3=a+0}-(\partial_3A_\mu)\vert_{x^3=a-0}=
\Pi_\mu^{\ \nu}A_\nu\vert_{x^3=a} \,.\label{match}
\end{eqnarray}
At this stage, one can forget the origin of $\Pi_\mu^{\ \nu}$ and quantize
the electromagnetic field subject to the conditions (\ref{match}) at
$x^3=a$ and to the conditions (\ref{condbc}) at $x^3=0$. This can be done,
at least at a somewhat formal level, and even the renormalization theory
can be developed, e.g. along the lines of Ref.~\cite{renorm}. The Casimir energy
density can be defined as a sum over the eigenfrequencies which then
are expressed trough the scattering data of the electromagnetic field.
There are several versions of this procedure leading
to different representations for the Casimir energy -- for details see Ref.~\cite{Casimir1}.

On the other hand, the Lifshitz approach \cite{Lifshitz} relates the Casimir energy density for
two parallel dielectric slabs to the corresponding
dielectric permittivities taken at the imaginary frequency. In some later works
\cite{Bordag:1995jz,Reynaud} the connection between two approaches was established and
generalization of the Lifshitz formula was presented. For the
interaction between two plane parallel interfaces separated by the distance $a$ and
possessing arbitrary reflection coefficients $r^{(1)}_{\rm TE, TM}$, $r^{(2)}_{\rm TE, TM}$
of the TE and TM electromagnetic modes on each of the surfaces it reads
\begin{equation}
    {\mathcal{E}}_L
    =\int\frac{d^3p_E}{16\pi^3} \ln [(1-e^{-2p_\| a}r_{\rm TE}^{(1)}r_{\rm TE}^{(2)})
        (1-e^{-2p_\| a}r_{\rm TM}^{(1)}r_{\rm TM}^{(2)})] .
        \label{EL}
\end{equation}
The reflection coefficients
are to be found from corresponding boundary or matching
conditions. For graphene with help of (\ref{match}) we obtain
at the Euclidean momenta
\be
    r_{\rm TE}^{(1)}=\frac {-\alpha \Phi}{2p_\| +\alpha\Phi}\,,\quad
    r_{\rm TM}^{(1)}=\frac {\alpha p_\| \Phi}{2\tilde p_\|^2 + \alpha p_\| \Phi}
    \label{rTETM},
\ee
while for the perfect conductor one has
\be
    r_{\rm TE}^{(2)}=-1,\quad
    r_{\rm TM}^{(2)}=1.
    \label{rTETM-c}
\ee
It is clear, that $\Phi$ must be rotated to Euclidean momenta as well.
We also note that the perfect conductor case is recovered
from (\ref{rTETM}) in the formal limit $\Phi\to\infty$.

The Euclidean momenta representation of the Casimir energy like (\ref{EL})--(\ref{rTETM-c}),
has several advantages. First of all,
it takes into
account contributions from possible surface plasmon modes not requiring
to analyze such modes explicitly \cite{Bordag:2005by}. Secondly, it is straightforward to consider its limiting cases
and also perform numerical evaluations as presented in the next section.

One can show by a direct computation that the energy $\mathcal{E}_1$, Eq.\ (\ref{E11}),
coincides with the leading $\alpha^1$ order in a perturbative expansion of the Lifshitz
formula (\ref{EL})-(\ref{rTETM-c}), so that the two approaches are consistent. In fact, the Lifshitz
formula is the one-loop vacuum energy (one closed vacuum loop) in an effective
theory corresponding to the action $S_{\rm M}+S_{\rm eff}$. As we have explained
above, the presence of the singular part $S_{\rm eff}$ is equivalent to the matching
conditions (\ref{match}). Imposing these conditions, in turn, is equivalent to summing
up the photon propagators with an arbitrary number of $\Pi^{mn}$ insertions (or, with
an arbitrary number of the fermion loop insertions). Therefore, the Lifshitz approach
corresponds to a partial summation of diagrams of the QFT approach. We shall
explain this correspondence in detail elsewhere.

\section{Results and discussion}
The formulae (\ref{E11}) and (\ref{EL})-(\ref{rTETM-c}) are suitable for the
numerical and asymptotical evaluation.
First we consider the
large separation limit, $a\to\infty$. Introducing dimensionless variables
$p_E\to p_E a$ in (\ref{EL})-(\ref{rTETM-c}) we are able to expand the integrand in a power series in $1/a$.
Each term of such expansion is integrable and in the leading order we get for the energy
\be\label{EL2}
{\mathcal{E}}_L\raisebox{-5pt}{$\sim\atop a\to \infty$}-\frac{\al N}{96\pi^2}\,\frac{2+v_f^2}{m a^4}\,.
\ee
We note that the energy is decreasing by one power of the separation $a$ faster than for ideal conductors.
Also we point out that this asymptotic expression is of
the first order in the coupling constant $\al$.

In the limit of small separation, $a\to0$, due to
the structure of the function $\Phi(p_E/a)$ the energy
\be\label{EL3}
{\mathcal{E}}_L\raisebox{-5pt}{$\sim\atop a\to 0$}
    \frac{1}{16 \pi a^3}\,h(\al,N,v_F)
\ee
factorizes into a distance dependent part and a function
$h(\al,N,v_F)$
independent of the separation $a$ but containing all powers of $\al$; in the leading order it is
\bea
\label{h1} &&
h(\al,N,v_F)= -\frac{N \al}{16} \times\\
&&\times
\left(1+\frac{2+v_F^2}{\sqrt{1-v_F^2}}\,{\rm arcsinh}\left(\frac{\sqrt{1-v_F^2}}{v_F}\right)\right)
+O(\al^2).\nonumber
 \eea

Therefore we see that the asymptotic behavior of the Casimir energy
in our model shows some surprising features being drastically different from that in the hydrodynamic model \cite{BV}--\cite{BGKM}.
At large separation it does not turn into the ideal conductor case,
while at small ones this case is regained.
Such behavior is counter-intuitive since the main contribution at small separations shall come from the high momenta for which one would expect the graphene film to become transparent. On the other hand, the behavior at large separation results from low momentum contributions for which the Dirac model is considered to be well proved
following directly from the electronic structure of graphene.

One can show that in the case of a massless
(gapless) fermions the same distance dependence as for ideal conductors
is retained for all distances due to lack of any dimensional
parameters in this case. The magnitude of the energy is defined
by the same asymptotic (\ref{EL3}) as for the $a\to0$ case.

\begin{figure}
\includegraphics[width=3.2in]{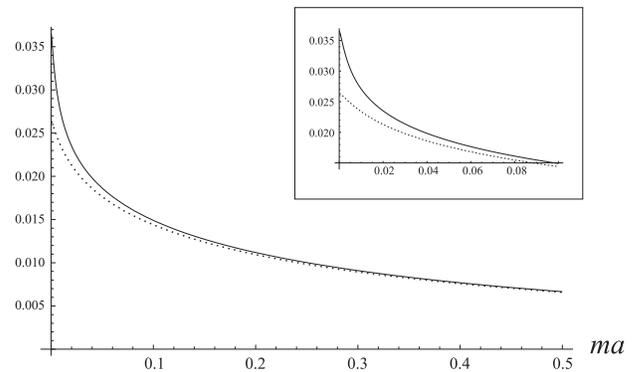}
\caption{\label{fig1}
The relative Casimir energy densities $\mathcal{E}_1/\mathcal{E}_C$ (solid line) and
$\mathcal{E}_L/\mathcal{E}_C$ (dashed line) as
functions of $ma$. A zoom of the small-distance region is
placed in the upper-right corner.}
\end{figure}

Let us now turn to the numerical evaluation. It is convenient to normalize the results to
the Casimir energy density
\begin{equation}
\mathcal{E}_C=-\frac{\pi^2}{720\, a^3} \label{ECas}
\end{equation}
for two plane ideal conductors separated by the same distance $a$. The relative quantities
$\mathcal{E}_1/\mathcal{E}_C$ and $\mathcal{E}_L/\mathcal{E}_C$ are dimensionless and depend
on a single dimensionless parameter $ma$.  To fix the scale, note that
for $m= 0.1$eV (actual values of $m$ are much smaller)
$ma=1$ corresponds to $a=1.97$ micrometer.
The results of calculations are depicted at Fig.\ \ref{fig1}.
For $m=0$ the
normalized energies $\mathcal{E}_{1,L}/\mathcal{E}_C$ are constant
independent of $a$ as explained above.

Thus, we can see that the magnitude of the considered Casimir interaction of
graphene with a perfect conductor is rather small.
Actual
measurement of such weak forces is a challenging, but by no means
hopeless, experimental problem
\footnote{
The linear dimensions
of existing suspended graphene samples reaches size of a couple of
millimeters (see Geim, Ref.~\cite{gra-rev})
which is well above the requirements of up-to-date Casimir
experiments \cite{Casimir1}. Also, the current precision of these
experiments is better than fractions of a percent, which is enough
to detect the Casimir interaction of graphene.}.
Strong dependence on the mass
parameter $m$ at large separation is also a characteristic feature
of the Casimir force. Getting an independent measurement of $m$
may be very important for our understanding of the electronic
properties of graphene. The mass of quasi-particles in graphene
is, probably, very tiny. This improves the detectability of the
Casimir interaction since the energy increases with decreasing
$m$.

As noted above,  the Casimir energy
$\mathcal{E}_1$ calculated in the QFT approach coincides   with
the lowest order expansion in the coupling constant $\al$ of the
Casimir energy $\mathcal{E}_L$ derived from the Lifshitz formula.
%
This
perturbative nature of the Casimir energy in graphene systems
(which is just the other side of its smallness) makes the
calculations much easier and will probably simplify the analysis
of other geometries like, e.g., folded or corrugated graphene near a
conducting surfaces. (An example of perturbative calculations
with $\delta$-potentials with the support of a non-trivial geometry
can be found in Ref.~\cite{Mpert}). This is in contrast with some other physical
effects in graphene which exhibit a strong coupling dependence.

There are several factors that can be included in the considered model to make it more realistic. Among them the effects of non-zero temperature, corrugation of the free-standing graphene sample, presence of impurities and non-vanishing density of carriers. However, for clean enough graphene samples, similar to Ref.~\cite{pureGr}, all mentioned factors can be considered as perturbations not changing the essentials of the Dirac model. In particular, the corrugation was shown to maintain the massless fermionic nature of the quasi-particles \cite{corrug}, and thus can be treated perturbatively as in Ref.~\cite{Mpert}. The effect of impurities can also be implemented into the theory as modifications to the Dirac operator which may be considered as perturbations at least in the sense of the operator theory. Indeed, while the impurities have a significant impact on the density of states \cite{Imp}
and broaden the Landau levels in the Hall regime, their effects may be represented through adding a complex chemical potential, expressed through
electronic self-energy \cite{Gusynin}. Although in general the frequency dependence of these terms can significantly modify the properties of the model, in many cases the constant approximation of such broadening of Landau levels works well \cite{constBroad} and
can be treated as corrections \cite{corrG}.

Thus, the method proposed above will therefore remain valid upon inclusion of these effects, and probably even numerically the corrections will remain small. Actual calculations will be the scope of our future work.

\begin{acknowledgments}
 This work was supported in parts by FAPESP (I.V.F.\ and D.M.G.), CNPq (D.M.G.\ and
D.V.V.), the grants RNP 2.1.1/1575 (I.V.F.\ and D.V.V.) and RFBR $07$--$01$--$00692$ (I.V.F.).\\
One of the
authors (M.B.) thanks the ESF Research Network CASIMIR for providing
excellent opportunities for discussion on the Casimir effect and
related topics.

\end{acknowledgments}

\end{document}